\documentclass[3p,times]{elsarticle}

\usepackage{ecrc}

\volume{00}

\firstpage{1}

\journalname{Nuclear Physics A}

\runauth{K.\,C.~Zapp}

\jid{nuphas}

\jnltitlelogo{Nuclear Physics A}

\usepackage{amsmath,amssymb}

\newcommand{\pt}{p_\perp}
\newcommand{\kt}{k_\perp}

\begin{document}

\vspace*{1cm}
\begin{flushright} 
IPPP/10/99\\ DCPT/10/198\\ MCnet/10/22 
\end{flushright}

\begin{frontmatter}


\title{Monte Carlo simulations of jet quenching in heavy ion collisions}

\author{Korinna Christine Zapp}

\address{Institute for Particle Physics Phenomenology, Durham University, Durham
DH1 3LE, UK}

\begin{abstract}
Jet quenching is one of the major discoveries of the heavy-ion program at
\textsc{Rhic}. While there is a wealth of data from \textsc{Rhic} that will
soon be supplemented with measurements at the \textsc{Lhc}, on the theoretical
side the situation is less clear. A thorough understanding of jet quenching is,
however, beneficial, as it is expected that medium-induced modifications of jets
allow one to characterise properties of the QCD matter produced in heavy ion 
collisions. This talk aims at summarising the main ideas
and concepts of the currently available Monte Carlo models for jet quenching. 
\end{abstract}


\end{frontmatter}

\section{Introduction}

The perhaps most general and most insightful definition of the Monte Carlo (MC)
method is given by Halton: \textit{``The Monte Carlo method is defined as
representing the solution of a problem as a parameter of a 
hypothetical population, and using a random sequence of numbers to construct a
sample of the population, 
from which statistical estimates of the parameter can be obtained.''}~\cite{
Halton:1970} This seems
to be rather unspecific, which is due to the fact that MC techniques have a very
broad range of applications to both statistical and deterministic problems. A
more pragmatic attempt is therefore made by James in defining: \textit{``A Monte
Carlo technique is any technique making use of random numbers to solve a
problem.''}~\cite{James:1980yn} While this is perhaps of more practical use, it
is still not very helpful for the understanding of how MC methods are used and
how they work. Narrowing the scope down to the applications under consideration
here, one can arrive at a more specific definition: In high energy physics, we
use our theories, physical intuition and models to devise computer 
programs that simulate high energy reactions by using random numbers to
construct synthetic events (Monte Carlo event generators).

\section{General considerations}

There are conceptually very different approaches to Monte Carlo generators. In
one extreme limit, the MC is a faithful representation of a theory. The parton
shower, which performs a resummation of large logarithms associated with
collinear gluon emission to leading logarithmic (LL) accuracy, falls into this
category. All parton shower implementations correctly reproduce the LL
behaviour, differences between the implementations (choice of evolution
variable, scale choices etc.) are beyond LL accuracy. In cases where a MC does
nothing else but represent a theory, it obviously has no tunable
parameters\footnote{Most MC generators tune the parton shower cut-off together
with the hadronisation, but the scale at which one stops trusting perturbation
theory is strictly speaking not a parameter of the theory or its MC
implementation. Also, some generators tune the scale of the strong coupling
constant in the parton shower, but this is again beyond LL.} and there is
control over the systematic accuracy.

In contrast to this, MC's can also be phenomenological models (in the extreme
case the model is defined by its MC implementation), that are used in cases
where there is little or no theoretical guidance. An example for this are
hadronisation models. This kind of MC's typically has parameters that have to
be tuned to data. While formal accuracy is not a meaningful concept in this
context, the quality of phenomenological models can be assessed using criteria
like accordance with physical principles, internal consistency or ability to
describe or predict a variety of experimental data.

\smallskip

Both approaches (and hybrids) are perfectly legitimate. Since MC generators
are used in very different ways, the choice of model or generator --
insofar as several exist -- will depend on the application. On the theory side,
one might want to use MC's to gain insight into the underlying mechanisms of
some phenomenon (this obviously applies to problems that cannot simply be
calculated in theory). In this case, a model that incorporates as many
theoretical constraints and as few parameters as possible is preferable. A
perfect description of experimental data does not have high priority,
a reasonable description of the qualitative behaviour is often sufficient.

The experiments, on the other hand, need MC tools to, for instance, estimate
the experimental efficiencies, corrections, backgrounds etc., or do unfolding.
For this kind of applications it is often advantageous to have models that are
more flexible and can be tuned to describe the data well. 

MC's are also used to extrapolate from (phase space) regions, where
measurements have been performed, to regions, where there is no data. In order
for the extrapolation to be reliable one needs a well constrained model with
not too many parameters. This may lead to a somewhat worse description of the
existing data than what more flexible and more phenomenological models can
achieve, but the predictive power is typically larger.

\medskip

MC generators have become indispensable tools in particle physics and all
important aspects of electron-positron and (anti)proton-proton collisions are
under control in the modern multi-purpose event generators
\textsc{Herwig}\cite{Bahr:2008pv}, \textsc{Pythia}\cite{Sjostrand:2007gs}
and \textsc{Sherpa}\cite{Gleisberg:2008ta}. Here, only the elements relevant for
jet production will be briefly introduced.
\begin{description}
 \item[matrix elements:] The hard matrix elements (ME) are calculated at fixed
order in perturbation theory and are available in the MC generators at leading
(LO) or next-to-leading (NLO) order.
 \item[final state parton shower:] The parton shower (PS) performs a resummation
of collinear logarithms to leading-logarithmic (LL) accuracy. It dresses all
(strongly interacting\footnote{Resummations of photon radiation also exist,
but will not be considered here.}) outgoing legs from the ME with collinear
radiation. At the infra-red cut-off scale the PS is interfaced with a
hadronisation model.
 \item[initial state parton shower:] The initial state PS exists only in
hadronic collisions and works like the final state PS, but has a boundary
condition at the infra-red cut-off scale that is given by the parton
distribution functions (pdf's). They describe the proton structure and are
obtained from global fits.
 \item[hadronisation:] Hadronisation is governed by non-perturbative physics
and so phenomenological models are employed.
\end{description}
This well established picture gets modified due to the presence of a dense
and strongly interaction medium when jet production in heavy-ion collisions is
considered. While a consistent theoretical treatment is still missing, general
considerations and observations lead to the following qualitative picture:
\begin{description}
 \item[matrix elements:] The ME's are not modified due to the high scale they
involve, which means that they are confined to very short length and time
scales and occur very early in the collision.
 \item[final state parton shower:] The final state PS evolves on a much longer
time scale and thus overlaps with the dense medium formed in the final state of
heavy-ion collisions. Modifications of the PS are thus expected, but no general
theory exists. There are, however, a number of calculations for special cases
like the single gluon radiation spectrum in the eikonal limit.
 \item[initial state parton shower:] The initial state PS evolves in only cold
nuclear matter density and was found to be unmodified at \textsc{Rhic} from
measuring direct hard photon production in nucleus-nucleus and high-$\pt$
hadron production in proton-nucleus collisions.
 \item[hadronisation:] Modifications of the hadronisation stage are generally
expected at least for hadrons at low and intermediate energies, but there is
no theoretical guidance.
\end{description}
Consequently, one has to rely on models for the description of medium modified
jets. Analytical calculations are successful in explaining leading hadron
suppression, but cannot be generalised to describe the complete jet
fragmentation pattern. Furthermore, no consistent quantitative picture has
emerged from studying leading hadrons. At \textsc{Rhic} reconstruction of jets
is under way and at the \textsc{Lhc} a large fraction of the jets will be
accessible above background. There is thus a need for MC generators modelling
jet quenching on the basis of multi-particle final states, as the experiments
need MC tools to perform their measurements
and there is a strong theoretical interest in studying sub-leading fragments.
The main reason is that sub-leading particles are likely to discriminate between
different microscopic mechanisms conjectured to underlie jet quenching. A
thorough understanding of medium-induced modifications of the entire jet
fragmentation pattern is thus requisite for a characterisation of the QCD matter
produced in heavy ion collisions based on jet quenching.
Furthermore, MC's may help to characterise also the jet-induced modifications
of the medium and to disentangle jets and background -- a problem that
receives attention from both the theoretical and experimental community.

\section{Update on JEWEL}

When one moves away from the eikonal limit one faces a couple of problems. For
instance, elastic and inelastic scattering are not well separated concepts any
more. Since (elastic) interactions with scattering centres in the medium can
transfer energy and momentum to the hard parton, they can induce radiation off
the projectile and thus become inelastic. Another problem is that 'vacuum' and
'medium induced radiation' are indistinguishable. These are probably the most
severe obstacles, although other issues like the radiation off the scattering
centre also have to be accounted for.

The solution to these problems implemented in the latest version of
\textsc{Jewel} is to describe scattering in the medium using $2\to 2$ partonic
matrix elements supplemented with parton showers as in proton-proton
collisions. In fact, in the standard description of hard proton-proton
collisions the parton shower is nothing but the bremsstrahlung of the hard
interaction and it therefore makes sense to use the same language for the
medium induced radiation. The advantages of this approach are that it naturally
generates elastic and inelastic interactions. The parton shower on top of $2\to
2$ ME's reproduces the collinear enhanced regions of the $2\to 2+n$ ME's and it
is known in principle how to systematically improve on this approximation.
Vacuum and medium induced radiation are not only incorporated in a common
framework, but have indeed become the same phenomenon and there is thus a
natural interplay between them. Finally, it is possible to include the
LPM-interference based on the probabilistic algorithm derived
in\cite{Zapp:2008af,lpm-long}.

\section{Overview over MC models}

It is the aim of this section to give a brief overview over the main ideas and
features of the existing MC models for jet quenching in roughly chronological
order, for details the reader is referred to the original publications.

\subsection*{HIJING}

\textsc{Hijing}\cite{Wang:1991hta,Deng:2010mv} simulates complete heavy-ion
events with hard and semi-hard (mini-)jet components and a soft part based on a
model for the formation and decay of colour strings. Hard matrix elements and
parton showers are simulated using \textsc{Pythia}.

\begin{description}
 \item[radiative energy loss:] Radiative energy loss is modelled by collinear
parton splittings, where the energy sharing is taken from an analytical
calculation. The colour connection between the two outgoing partons is assumed
to be broken so that the colour (string) topology of the event changes.
 \item[collisional energy loss:] There is no collisional energy loss.
 \item[medium:] The energy loss model uses a constant mean free path and
screening length.
 \item[hadronisation:] The Lund string model is used for hadronisation. The
softer parton distribution together with the modified colour topology leads to a
softer hadronic ensemble.
 \item[example for an application:] Due to its ability to generate complete
heavy-ion events, \textsc{Hijing} has for instance been used to study the
influence of background fluctuations on jet
measurements\cite{Alessandro:2006yt}.
\end{description}

\subsection*{PYQUEN/HYDJET}

\textsc{Hydjet}\cite{Lokhtin:2008xi} also simulates complete heavy-ion events,
but in contrast to \textsc{Hijing} it uses a hydrodynamical model for the soft
component. Jet production and interactions of the jets in the medium are
modelled using \textsc{Pyquen}\cite{Lokhtin:2005px}, which again is built on
matrix elements and parton showers of \textsc{Pythia}.

\begin{description}
 \item[radiative energy loss:] At each scattering a gluon sampled from a
BDMPS radiation spectrum is radiated under a finite angle. The radiated gluon
stays within the colour singlet system of the parent parton.
 \item[collisional energy loss:] The collisional energy loss is based on a
high-energy approximation of the LO perturbative $2\to 2$ partonic scattering
matrix elements. 
 \item[medium:] The medium is modelled as a quark-gluon fluid undergoing
boost-invariant longitudinal expansion (Bjorken model).
 \item[hadronisation:] Finally, the hard component of the event is hadronised
using the Lund string model. 
 \item[example for an application:] \textsc{Pyquen/Hydjet} runs very fast and
has thus been used to extract medium modified fragmentation
functions\cite{D'Enterria:2007xr}.
\end{description}

\subsection*{JEWEL}

It is the objective of \textsc{Jewel}\cite{Zapp:2008gi,Zapp:2008af} to develop a
dynamically consistent MC model for jet quenching, that is consistent with all
analytically known limiting cases. It is the most specialised project in the
sense that it does not simulate matrix elements, but only the final state
parton shower and hadronisation. 

\begin{description}
 \item[radiative energy loss:] In its newest form \textsc{Jewel} has a pQCD
based microscopic model for interactions in the medium including a local
implementation of the LPM-effect.
 \item[collisional energy loss:] Collisional energy loss is naturally included
in the above model.
 \item[medium:] For the medium several options are foreseen: a semi-analytical
Bjorken model and an interface to hydrodynamic calculations.
 \item[hadronisation:] For hadronisation an adapted version of the Lund string
model (based on \textsc{Pythia}) is used.
 \item[example for an application:] As a consequence of the detailed and
democratic treatment of medium interactions and jet evolution \textsc{Jewel} is
well suited for jet observables like event shapes and jet
rates\cite{Zapp:2008gi}.
\end{description}

\subsection*{Q-PYTHIA/Q-HERWIG}

The philosophy of the
\textsc{Q-Pythia/Q-Herwig}\cite{Armesto:2009fj,Armesto:2009ab} project is to
construct a minimal, theory based MC model for medium-modified jets.
\textsc{Q-Pythia/Q-Herwig} simulate only jets, the matrix elements are taken
from \textsc{Pythia} and \textsc{Herwig}, respectively. The parton shower
routines are modified such that the final state parton shower has a modified
splitting function.

\begin{description}
 \item[radiative energy loss:] The splitting function in the final state parton
shower is written as $P_\text{tot}=P_\text{vac} + \Delta P$, where $\Delta P$
is derived from a BDMPS gluon spectrum under the assumption that the relation
of gluon spectrum and splitting function is the same as for vacuum emissions.
The treatment of colour is the same as in vacuum.
 \item[collisional energy loss:] There is no collisional energy loss.
 \item[medium:] The medium has to be specified by the user.
 \item[hadronisation:] \textsc{Q-Pythia} and \textsc{Q-Herwig} use the Lund
string model and \textsc{Herwig}'s cluster hadronisation, respectively. 
 \item[example for an application:] Because of its consistent treatment of
intra-jet dynamics \textsc{Q-Pythia} has for instance been used to investigate
jet broadening\cite{Abeysekara:2010ze}.
\end{description}

\subsection*{YaJEM}

YaJEM\cite{Renk:2008pp,Renk:2009nz} also modifies the \textsc{Pythia} parton
shower. It is assumed that the dominant effect of interactions in the medium is
that the virtuality of hard and semi-hard partons is increased.

\begin{description}
 \item[radiative energy loss:] The virtuality of partons is increased between
the splittings generated by the \textsc{Pythia} parton shower by an amount
given by a local transport coefficient.
 \item[collisional energy loss:] The energy and momentum of partons is reduced
by an amount given by a local drag coefficient.
 \item[medium:] The medium is taken from a hydrodynamic calculation.
 \item[hadronisation:] Hadronisation is taken care of by the Lund string model
in \textsc{Pythia}, the colour topology is assumed to be the same as in vacuum.
 \item[example for an application:] YaJEM can be used to study jet observables
like the angular distribution of partons forming a jet\cite{Renk:2008pp}.
\end{description}

\subsection*{MARTINI}

\textsc{Martini}\cite{Schenke:2009gb} is built on the \textsc{Amy} results for
partonic energy loss. It also uses \textsc{Pythia} for matrix elements, parton
shower generation and hadronisation.

\begin{description}
 \item[radiative energy loss:] After the final state parton shower the partons
are propagated through the medium with the help of rate equations, where the
transition rates are taken from \textsc{Amy}. There is no exchange of colour
with the medium so that radiated gluons remain colour correlated with the parent
parton.
 \item[collisional energy loss:] Transition rates for elastic scattering are
calculated in pQCD and included in the same way as the inelastic processes.
 \item[medium:] Different hydrodynamic calculations can be used.
 \item[hadronisation:] Hadronisation is performed using the Lund string model.
 \item[example for an application:] Nuclear modification factors have been
calculated with \textsc{Martini}\cite{Schenke:2009gb}.
\end{description}

\section{Summary and Outlook}

The MC landscape is clearly diverse and cannot easily be summarised. Still,
there are some noticeable trends.
\begin{itemize}
 \item Most of the models use hydrodynamical calculations as model for the
medium. There is, however, so far no systematic investigation of the
sensitivity of classes of observables to details of the modelling of the
medium. 
 \item The majority of models implements a democratic treatment of all partons.
This is very important in order to arrive at a consistent modelling of
sub-leading fragments and has potentially large impact on observables.
 \item Some models consider modifications of the parton shower based on the
argument that the parton shower evolves on a time scale that is comparable to
the lifetime of the medium and the transverse size of the nuclear overlap
region.
 \item In line with the argument that radiative processes are the dominant
source of energy loss for energetic partons all MC's include a model for
radiative energy loss. 
 \item It was found that collisional energy loss can be numerically important
and by now it is included in most of the models.
 \item Induced radiation and the LPM suppression are mostly treated in an
effective way owing to the difficulties associated with the treatment of
quantum  mechanical interference effects in probabilistic frameworks such as
MC's.
 \item There is a hadronisation monoculture in the sense that except for
\textsc{Q-Herwig} all projects use the Lund string model. The different
treatment of colour underlines a fundamental problem, namely that
hadronisation in a nuclear environment is poorly understood.
\end{itemize}

\smallskip

Instead of aiming for a comprehensive review of all recent developments, only a
few shall be exemplarily highlighted here. 

A first study of induced radiation off a colour dipole was carried out and
found an angular separation of medium-induced and vacuum
radiation\cite{MehtarTani:2010ma}.

In a detailed investigation the probabilistic formulation of the
LPM-effect\cite{Zapp:2008af} was shown to quantitatively reproduce the BDMPS
results\cite{lpm-long}.

First attempts have been made to characterise the effect of jet energy loss on
the medium in the context of a hydrodynamical calculation\cite{Neufeld:2010tz}
and in a particle picture\cite{Zapp:2008gi}.

\smallskip

There remain, obviously, a number of issues that have to be addressed by future
investigations. The interplay of vacuum and medium-induced emissions in
realistic kinematics, for instance, is largely an unsolved problem. There is
some guidance
from analytical calculations in near-eikonal kinematics, but generalisation to
realistic scenarios is far from trivial. 

Another problem is uncertainties. As discussed earlier, phenomenological models
have no formal accuracy. However, uncertainties can sometimes be quantified by
mapping out a range of reasonable models or assumptions. The dependence on the
choice of evolution variable could for instance by studied by comparing results
from angular, virtuality and $\kt$ ordered parton showers. Large uncertainties
can -- depending on the observable -- also be associated to the exact modelling
of hadronisation. Here, using different models and comparing different ideas of
how colour should be treated could also help to get an estimate of the size
and systematics of uncertainties.

Jet-induced modifications of the medium are not yet well understood but are
potentially important for jet measurements at the \textsc{Lhc}. Large
background contributions have to be subtracted from the total energy measured
inside a jet to arrive at the true jet energy. The background is estimated in
regions outside the jet so that a significant modification of the background
close to the jet may bias the extracted jet energies. 

\medskip

The description of jets in heavy-ion collisions is largely based on models, as 
a universally valid theory is missing. As there is no consensus about the
microscopic mechanisms responsible for interactions of hard partons in a dense
medium and thus the phenomenon of jet quenching or the language that should be
used to describe them a variety of (Monte Carlo) models have been developed.
In this situation the application dictates the choice of the tool. If used
wisely MC models in conjunction with \textsc{Lhc} jet data have the
potential to play a central role in deciphering jet quenching.

\bibliographystyle{elsarticle-num}
\bibliography{hp10-refs}

\end{document}